\documentstyle[psfig,referee]{mn}

\title[Multiple delays in QSO 0957+561]{Multiple delays in QSO 0957+561:
observational evidence and interpretation}
\author[L.J. Goicoechea]
       {L.J. Goicoechea \\
	Departamento de F\'{\i}sica Moderna, Universidad de Cantabria,
	Avda. Los Castros s/n, E-39005 Santander, Spain \\ 
	E-mail: goicol@unican.es}
	
\begin{document}

\maketitle

\begin{abstract}
Q0957+561A,B is a double-imaged quasar that has been intensively observed 
during the last 10 years in different optical bands and with several 
telescopes, and we concentrated on recent public data obtained at the Apache 
Point Observatory (APO) and the Teide Observatory (TO). When an intrinsic event
appears in the light curve of Q0957+561A, its twin event (a similar feature) 
is seen in the brightness record of Q0957+561B, and thus, one can measure the 
corresponding time delay. The TO dataset includes two prominent twin events, 
which were detected with a time separation of 425 $\pm$ 4 days. On the other 
hand, from the APO dataset, we found a clear evidence for two different time 
delays associated with two pairs of twin events: 417.0 $\pm$ 0.6 (APO main twin
events) and 432.0 $\pm$ 1.9 days (APO secondary twin events), where the 
APO(main)--APO(secondary) difference delay is of $-$ 15 $\pm$ 2 days. In 
agreement with the Yonehara's idea, if the three pairs of twin events are 
originated inside a standard hybrid source (accretion disk and circumnuclear 
stellar region), the three measured time delays indicate that they do not come 
from a common zone in the source. Therefore, we can consider that the prominent
features are caused by flares in a standard hybrid source and discuss on the 
size and nature of the region of flares. In this paper it is showed that the 
more plausible interpretation is that two of the three flares are generated at 
distances (from the central black hole) larger than 90 pc. Some stellar 
scenarios can explain the two flares far away from the black hole, while 
phenomena in the accretion disk cannot cause them.  
\end{abstract}


\section{Introduction}

The observed optical luminosity of QSOs is probably generated in a source 
including an accretion disk around a supermassive black hole and a 
circumnuclear stellar region (e.g., Umemura, Fukue \& Mineshige 1997, 
and references therein). Therefore, some
kind of intrinsic variability could be due to local and violent 
physical phenomena (flares) taking place in the accretion disk (e. g.,
hot-spots) or in the innermost stellar region (e. g., star bursts).
These two standard regions for flares have very different sizes, and
thus, if we were able to map several locations of flares within a
source QSO, we would measure the size of the RF (region of flares) and
determine the environment in which the local and violent variability
is originated (Yonehara 1999). A standard accretion disk has a radius 
of the order of 10$^{-2}$ pc, whereas the innermost stellar region may
be extended as far as 10$^2$-10$^3$ pc.  

In a pioneer work, Yonehara (1999) has suggested a way to map the
positions of the intrinsic flares in a gravitationally lensed quasar.
For a double-imaged QSO, with optical images A and B, there is a time 
delay between a given intrinsic event in image A and its twin event
(a similar feature) in image B. However, if the optical variability is
generated in different points of a source with finite size, it must be
not expected a unique time delay for different pairs of twin events.
So, the discovery of multiple delays in a double QSO indicates the
presence of twin events induced by flares, and the time delay 
distribution must inform us about the size and nature of the RF. In 
this paper (Section 2), we searched for multiple delays in the light 
curves of the lensed quasar Q0957+561A,B. The light curves of both 
images in this gravitational mirage show variability on very 
different timescales, but we concentrated on the well-sampled intrinsic 
events with an amplitude of about 100 mmag and lasting several months 
($\sim$ 100-300 days), which were recently found by Kundi\'c et al. 
(1995, 1997) and Serra-Ricart et al. (1999). 

In Section 3, we interpret the results derived from the careful 
analysis of the best available optical light curves of Q0957+561A,B. 
Finally, we summarize our conclusions in Section 4. Scenarios for 
flares as diverse as accretion disk instabilities, disruption of 
stars in the gravitational field of a supermassive black hole, stellar 
collisions or violent stellar evolution are also put in perspective in this 
last section (see Cid Fernandes, Sodr\'e \& da Silva 2000, and references 
therein). 

\section{Time delay(s) between the components A and B of QSO 0957+561}

\subsection{Historical background and description of the data}

The two optical components of QSO 0957+561 have been monitored during 
about 20 years (Lloyd 1981; Keel 1982; Florentin-Nielsen 1984; 
Schild \& Cholfin 1986; Vanderriest et al. 1989; Schild 1990; Schild
\& Thomson 1995; Kundi\'c et al. 1995; 1997; Serra-Ricart et al. 1999;
Oscoz et al. 2001; Slavcheva-Mihova, Oknyanskij \& Mihov 2001), and the 
datasets inferred from the observations have been usually analyzed from
standard techniques, which look for a unique time delay between Q0957+561A
and Q0957+561B. However, only if all features come from a common zone in the
source (or the emitting points are included within a very small region),
we can properly speak of a well-defined time delay. The use of a standard 
procedure, in general, leads to an effective time delay that will correspond 
to either the time delay associated with the dominant twin features or 
the average of several time delays associated with several pairs of twin 
features in the light curves of the system.   

The best available optical light curves, in terms of sampling rate, 
signal-to-noise ratio and intrinsic character, were obtained by Kundi\'c 
et al. (1995, 1997) and Serra-Ricart et al. (1999), and these trends could
shed light on the existence of a well-defined delay (when there are no 
discrepancies between time delays for different pairs of twin events or
the possible discrepancies cannot be resolved) or the opposite case. Thus,
we concentrate our attention on the events with a width of $\sim$ 100-300
days and an amplitude of about 100 mmag (sharp features with high 
signal-to-noise ratio), which are included in the brightness record obtained 
at the Apache Point Observatory (APO) and the Teide Observatory (TO) for the
period 1995-1998. Another large dataset compiled by R. Schild shows clear 
evidences in favor of extrinsic variability (Schild 1996), and therefore, 
we avoid this dataset. We remark that Schmidt \& Wambsganss (1998) and 
Gil-Merino et al. (2001) have not found reliable microlensing imprints in 
the records of brightness by Kundi\'c et al. (1995, 1997) and Serra-Ricart 
et al. (1999), respectively, although Schmidt \& Wambsganss (1998) showed
a difference light curve with an {\it anomaly} during the central dates which
is directly related with the subject of this paper (a group of data is 
slightly but coherently above the zero-line and the next group of data 
is coherently below the zero-line). 

\subsection{First evidence for multiple delays}
 
The observations carried out at TO from 1996 through 1998 (in the $R$ band) 
were used to compare the light curves of images A and B and obtain 
effective time delays using pairs of seasons (Serra-Ricart et al. 1999). 
In principle, two comparisons are possible: 1996(A)-97(B) seasons and 
1997(A)-98(B) seasons, but the signal-to-noise ratio in the first pair was 
very small, and only the two twin features seen in the last pair (with an 
amplitude of 120 mmag and a duration of about 300 days; see Serra-Ricart 
et al. 1999; Gil-Merino et al. 2001) can be seriously considered to measure 
an accurate time delay. Using the data included in this last pair 
of seasons and the $\delta^2$-test, Serra-Ricart et al. (1999) derived a 
reliable delay of 425 $\pm$ 4 days (1 $\sigma$). As they basically compared 
two twin events, we do not consider their detection as an effective delay but 
the time delay between the quoted prominent features. So, taking into 
account the TO events with $S/N \sim$ 2.5 (we define the signal-to-noise 
ratio as the ratio between the semi-amplitude of the events and the mean 
photometric error), it is inferred an optimal delay of 425 days. We remark that
brightness records with $S/N \sim$ 1 are not suitable for measuring time
delays, because the techniques fail in this situation. With regard to this
issue, the section 3 of Pijpers (1997) is very instructive. In order to infer
reliable delays, it seems reasonable to work with signals characterized by 
$S/N >$ 2.

\begin{figure}
\psfig{figure=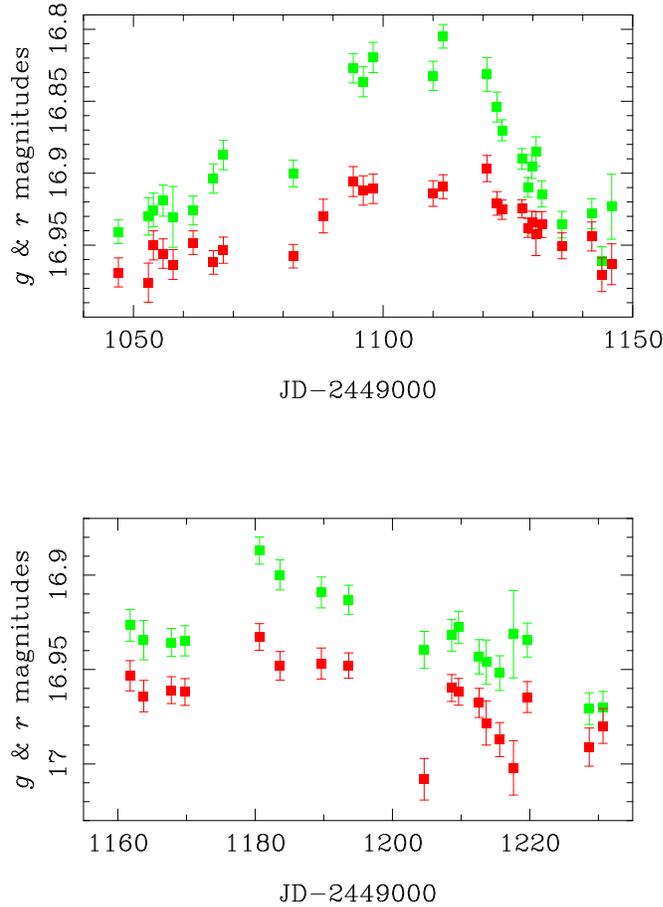,width=10cm}  
\caption{The main event (top panel) and the secondary one (bottom panel) seen
in the light curve of Q0957+561B obtained at the Apache Point Observatory
during the 1996 season. The light filled squares represent the photometric 
data in the $g$ band, while the dark filled squares correspond to the data in 
the $r$ band.}
\label{Fig. 1}
\end{figure}

\begin{figure}
\psfig{figure=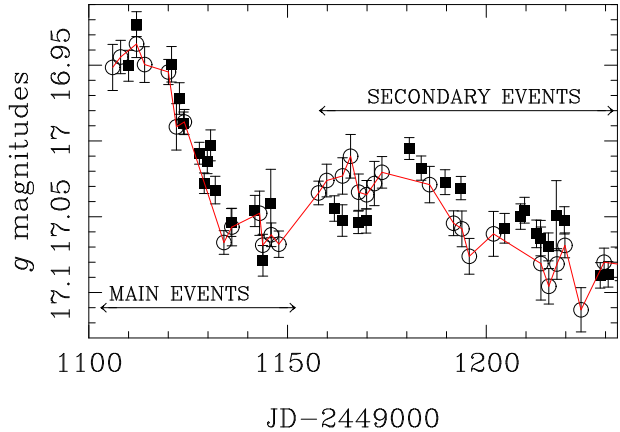,width=10cm,angle=-90}  
\caption{Combined photometry of Q0957+561A,B for the 1995/1996 seasons
in the $g$ band (at Apache Point Observatory). The open circles trace the 
time-shifted (+ 417 days) $A95$ and the filled squares trace the
magnitude-shifted (+ 0.118 mag) $B96$. The line defines the
shifted brightness record of the image A. An {\it anomaly} appears from day 
1160 to day 1230: firstly, the circles are coherently above the squares, 
and after, the opposite case occurs.}
\label{Fig. 2}
\end{figure}

\begin{figure}
\psfig{figure=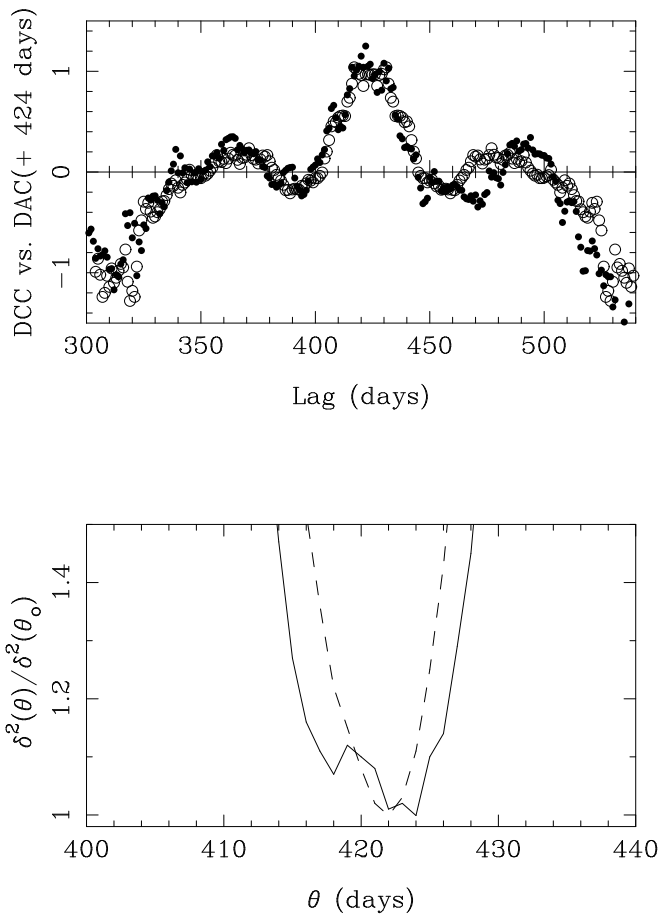,width=10cm}  
\caption{The $\delta^2$-test from APO data in the $g$ band. In the top panel we
compare the discrete cross-correlation function ($DCC$; filled circles) and 
the discrete autocorrelation function ($DAC$) shifted by 424 days (open
circles). We use bins with size of 5 days. In the bottom panel it showed the
normalized $\delta^2$ for two different time resolutions: bins with size of 5
days (solid line) and bins with size of 15 days (dashed line).}
\label{Fig. 3}
\end{figure}

\begin{figure}
\psfig{figure=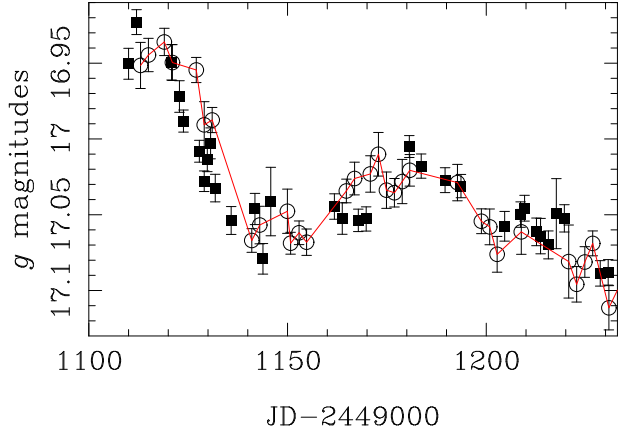,width=10cm,angle=-90}  
\caption{Combined photometry of Q0957+561A,B for the 1995/1996 seasons
in the $g$ band (at Apache Point Observatory). $A95$ is now time-shifted by 424
days, in agreement with the result of the $\delta^2$-test (bins with size of 5
days).}
\label{Fig. 4}
\end{figure}

\begin{figure}
\psfig{figure=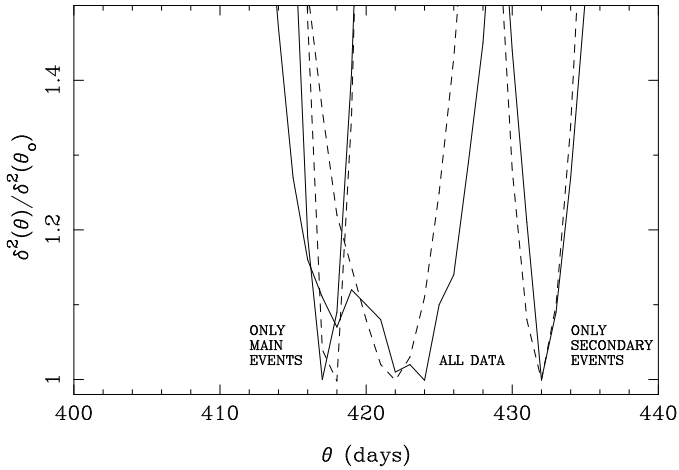,width=10cm,angle=-90}  
\caption{Normalized $\delta^2$ from all APO data (see Fig. 3), only APO data
in the main events and only APO data in the secondary events. The results for 
bins with size of 5 days (solid lines) and bins with size of 15 days (dashed 
lines) are drawn. We note the {\it breaking} of the relatively broad central
peak (all data) in two narrow peaks that don't overlap. It is evident that 
two different delays are needful.}
\label{Fig. 5}
\end{figure}

\begin{figure}
\psfig{figure=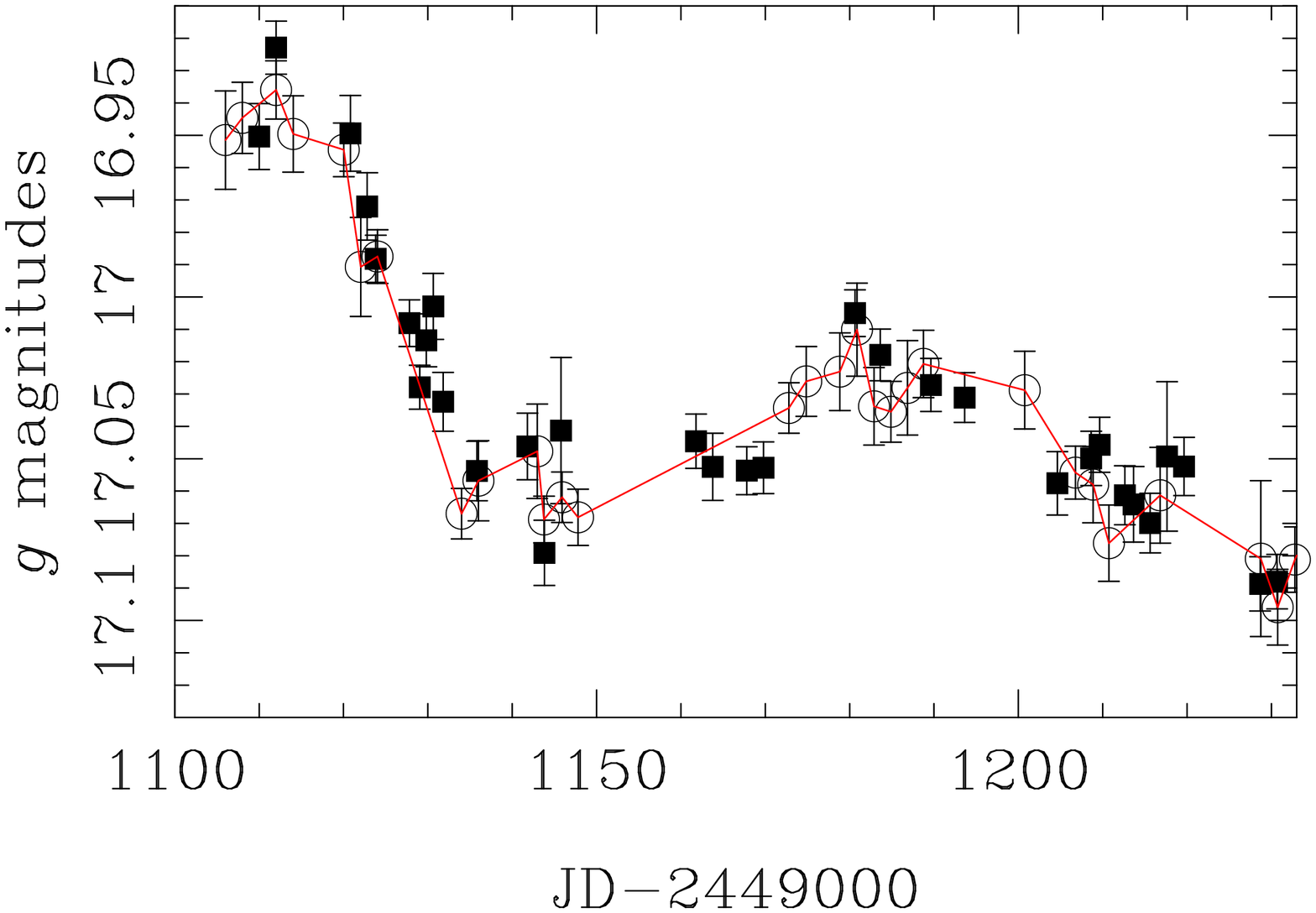,width=10cm,angle=-90}  
\caption{Combined photometry of Q0957+561A,B for the 1995/1996 seasons
in the $g$ band (at Apache Point Observatory). The first part of $A95$ is 
time-shifted by 417 days (main drop) and the second part is time-shifted by 432
days (secondary event). No {\it anomaly} is present.}
\label{Fig. 6}
\end{figure}

On the other hand, CCD images of Q0957+561A,B were taken with the APO 3.5 m 
telescope in the $g$ and $r$ bands, during the 1995 and 1996 seasons
(Kundi\'c et al. 1995, 1997). The light curve of the image A in the first
season ($A95$) exhibited a significant variability. In particular, $A95$
included a sharp drop of about 100 mmag in 1994 December (data in $A95$
encompass 6 months of observations between 1994 December 2 and 1995 May 31),
which represents the second half of a whole event. The twin event of this
feature in $A95$ was observed in the light curve of the image B during the
1996 season ($B96$). In the $g$ band, the twin event is characterized by
an amplitude and a width of 130 mmag and about 100 days, respectively. 
Just after these two APO main events in $A95$ and $B96$, we can find other
two sharp twin features. The APO secondary events have a duration slightly
less than the main ones and an amplitude of 50-70 mmag in the $g$ band.
Fig. 1 shows the main event (top panel) and the secondary event (bottom 
panel) in $B96$. The data in the $g$ band (light filled squares) and in
the $r$ band (dark filled squares) are depicted. We note the high 
signal-to-noise ratios in the $g$ band: $S/N \sim$ 6.5 in the main event
and $S/N \sim$ 3 in the secondary one. The $(S/N)_{g-band}$ value in the 
secondary events is similar to the $(S/N)_{r-band}$ value in the main 
features. The secondary events in the $r$ band are very noisy. Using 
$A95$ and $B96$ in the $g$ band, and three different techniques to infer
the effective time delay (linear interpolation, optimal reconstruction and 
dispersion spectrum), Kundi\'c et al. (1997) derived a best value of 417 
days. Fig. 2 displays together $A95$ (open circles) and $B96$ (filled
squares) in the $g$ band. $A95$ was shifted by a time delay of 417 days,
while $B96$ was shifted by an optimal magnitude offset of 118 mmag (see
Kundi\'c et al. 1997). Only the main overlaping region of $A95$ and $B96$
is showed, i.e., that one including the two pairs of twin features, and to
guide the eye, a line joins the data for the A component. One sees in
Fig. 2 an excellent agreement between the APO main twin features 
(from day 1100 to day 1150) as well as an $anomaly$ with regard the two 
APO secondary events. The secondary event in the flux of image A is not 
aligned with the secondary one in the flux of image B, or in other words,
the secondary event in the time-shifted (+ 417 days) $A95$ seems still 
delayed with respect to its twin event in $B96$. As an additional problem,
the best solution for the delay from TO data (425 days) is out
of the 95 per cent confidence interval claimed by Kundi\'c et al. (1997):
417 $\pm$ 3 days. From another point of view, the solution of 417 
days is in clear disagreement with the TO photometry (see Fig. 16 in 
Serra-Ricart et al. 1999).

The three methods used by Kundi\'c et al. (1997) to infer an optimal 
effective time delay of 417 days, seem very sensitive to the two main 
features. In a manner of speaking, these classical methods forget the
rest of data, and we need to find some discrete technique to compare
all information in both brightness records. The discrete cross-correlation
method has several variants which may be sensitive to the whole light
curves, and we choose the $\delta^2$-test (e.g., Serra-Ricart et al.
1999), in which the shifted discrete autocorrelation function ($DAC$) is
matched to the discrete cross-correlation function ($DCC$). From the
$\delta^2$-test we infer optimal effective time delays of 424 days (using
bins with size of 5 days) and 422 days (using bins with size of 15 days),
both in agreement with the measurement from TO data. We note that the TO data
were studied with a relatively poor time resolution (bins with size of 
40 days), while we analyzed the APO data with a good time resolution:
smoothing size of 5--15 days. In Fig. 3 (top panel) the $DAC$ for the A 
component shifted by 424 days (open circles) and the $DCC$ (filled circles) 
are presented. In this $DAC$ vs. $DCC$ comparison we used bins with size of 
5 days (results with the best time resolution). Using bins with size of 5 
days (solid line) and 15 days (dashed line), possible values of the 
effective time delay ($\theta$) versus the associated $\delta^2(\theta)$ 
values, normalized by its minimum value $\delta^2(\theta_0)$, have been also 
plotted in Fig. 3 (bottom panel). As it was already mentioned, the new optimal
values close to 425 days are in good agreement with the measurement made
by the IAC group. However, is there a good alignment between twin 
features?. To answer this question we have redrawn in Fig. 4 the combined
photometry of Q0957+561A,B for the 1995-96 seasons in the $g$ band, but
shifting $A95$ by a time delay of 424 days. We conclude that an effective 
delay of 424 days is not able to align neither the APO main events nor 
the APO secondary events, although this larger value can be considered as 
the estimate of an average time delay. The 424 days solution leads to some
discrepancies whose imprints can be seen in the top panel of Fig. 3 (e.g., 
there is a clear lag between the "humps" to the right of the central peaks).
{\it Only the existence of two different delays can lead to the alignment of 
both pairs of twin events}.

\subsection{Detection of multiple delays}

Our last task must be to obtain, in a consistent way, the time delay 
between the APO main events and the time delay between the APO secondary
events. From the data included in the sharp drop of the main events, we make
a first $\delta^2$ function. Moreover, using the data corresponding to 
the secondary events, we do a second $\delta^2$-test. The two new
$\delta^2$ functions together with the $\delta^2$ function inferred from
all data (see bottom panel of Fig. 3) are showed in Fig. 5. When we only 
take data corresponding to the main events, we obtain a narrow peak in
$\delta^2$  and recover the expected result of a delay of 417 days, 
whereas if only data in the secondary events are used, then another narrow
peak centered on 432 days is derived. In Fig. 6, we draw our best
solution for the 1995-96 seasons in the $g$ band: the first part of 
$A95$ is shifted by + 417 days (main drop) and the second part is 
shifted by + 432 days (secondary event). Finally, to sum 
up, a time delay of 425 $\pm$ 4 days explains the signal recorded at Teide 
Observatory in the $R$ band (two twin events with a relatively long 
time-scale of about 300 days), while a unique time delay does not align
the two pairs of twin events detected at Apache Point Observatory in the
$g$ band. From the APO light curves, a double time delay of 417.0 $\pm$ 0.6 
days (for the main features) and 432.0 $\pm$ 1.9 days (for the secondary 
events) is favored, and thus, {\it three different delays are detected from
APO/TO data} (the APO/Main-APO/Secondary difference delay is of $-$ 15.0 $\pm$
2.0 days, the APO/Main-TO difference delay is of $-$ 8.0 $\pm$ 4.0 days and
the APO/Secondary-TO difference delay is of $+$ 7.0 $\pm$ 4.4 days; see
Table 1). 

\begin{table}
 \centering
 \caption{Time delays from APO data.}
 \begin{tabular}{@{}lcccc}
  Events    & $S/N$ & Delay (days) & APO-APO difference delay (days) 
            & APO-TO difference delay (days) \\
  Main	    & $\sim$ 6.5 & 417.0 $\pm$ 0.6  & $-$ 15.0 $\pm$ 2.0
            & $-$ 8.0 $\pm$ 4.0 \\
  Secondary & $\sim$ 3.0 & 432.0 $\pm$ 1.9  & $+$ 15.0 $\pm$ 2.0
            & $+$ 7.0 $\pm$ 4.4 \\
 \end{tabular}
\end{table}

We computed 1 $\sigma$ uncertainties (68.3 per cent confidence limits) for the 
time delays inferred from the $\delta^2$-test. Each 1 $\sigma$ confidence
interval was derived through intensive simulations: we made $10^5$ possible
underlying signals which are consistent with the observed light curve of the
image A, and $10^5$ possible underlying brightness trends of the image B (in
agreement with the photometric behaviour of this image). From the $10^5$ pairs
of possible underlying signals, we can estimate $10^5$ time delays (using the
$\delta^2$-test), and therefore, the distribution of delays and the standard
confidence interval. The whole procedure is similar to the methodology used by 
Serra-Ricart et al. (1999). By means of intensive simulations, other studies 
are also interesting and viable. For example, from the APO main twin events, we
found that 99.9 per cent of the delays are included in the range 417 $\pm$ 2
days. On the other hand, from the APO secondary twin events, the delay
distribution is broader. However, a delay $\ge$ 425 days has a probability
higher than 99 per cent. So, we unambiguously found two different time delays
in the APO light curves.

\section{Interpretation of the results}

We assume that the observed optical luminosity of QSO 0957+561 is due to a
standard hybrid source (accretion disk and circumnuclear stellar region), and 
consequently, exotic sources are not taken into account in this paper (e.g., 
two accretion disks around two close supermassive black holes). The total 
luminosity is the superposition of a dominant non-variable "background" 
component and a variable part, where the background luminosity is mainly 
originated in either the full accretion disk or the full stellar environment, 
and the QSO variability can be made by flares or another kind of activity. In 
this scenario, the APO/TO events must be caused by flares in the source QSO.

If the source centre is placed at an angular position $\mbox{\boldmath 
$\beta$}$, the time delay between its images A (at 
$\mbox{\boldmath $\theta$}_A$) and B (at $\mbox{\boldmath $\theta$}_B$) is 
given by (e.g., Schneider, Ehlers \& Falco 1992)
\begin{equation}
\Delta\tau_{BA}(\mbox{\boldmath $\beta$}) = \frac{D}{2c} (1+z_d)
[\mbox{\boldmath $\theta$}_B^2 - \mbox{\boldmath $\theta$}_A^2 +
2(\mbox{\boldmath $\theta$}_A - \mbox{\boldmath $\theta$}_B).\mbox{\boldmath 
$\beta$} - 2\psi(\mbox{\boldmath $\theta$}_B) + 
2\psi(\mbox{\boldmath $\theta$}_A)] ,
\end{equation}
where $D$ = $D_sD_d$/$D_{ds}$ ($D_d$, $D_s$ and $D_{ds}$ are the angular
diameter distances from us to the lens, from us to the source and from the lens
to the source, respectively), $c$ is the velocity of light in vacuum, $z_d$ is
the redshift of the lens and $\psi$ is the deflection potential, which is
directly related to the scaled deflection angle $\mbox{\boldmath $\alpha$}$: 
$\mbox{\boldmath $\alpha$}$ = $\mbox{\boldmath $\nabla$}\psi$. However,
when a flare occurs, the corresponding signal originated at an angular position 
$\mbox{\boldmath $\beta$} + \delta\mbox{\boldmath $\beta$}$ will be
gravitationally lensed at $\mbox{\boldmath $\theta$}_A + 
\delta\mbox{\boldmath $\theta$}_A$ and 
$\mbox{\boldmath $\theta$}_B + \delta\mbox{\boldmath $\theta$}_B$. Therefore, 
from Eq. (1), considering an expansion up to the first
order in the angular displacements ($\delta\mbox{\boldmath $\beta$}$, 
$\delta\mbox{\boldmath $\theta$}_A$, $\delta\mbox{\boldmath $\theta$}_B$) and
the lens equation ($\mbox{\boldmath $\nabla$}\psi$ = 
$\mbox{\boldmath $\alpha$}$ = $\mbox{\boldmath $\theta$} - 
\mbox{\boldmath $\beta$}$), we infer the time delay difference
\begin{equation}
\delta(\Delta\tau_{BA}) = \Delta\tau_{BA}(\mbox{\boldmath $\beta$} + 
\delta\mbox{\boldmath $\beta$}) - \Delta\tau_{BA}(\mbox{\boldmath $\beta$})
= \frac{D}{c} (1+z_d) [(\mbox{\boldmath $\theta$}_A - 
\mbox{\boldmath $\theta$}_B).\delta\mbox{\boldmath $\beta$}].
\end{equation}
Using the distance displacement in the source plane $\delta{\bf r}$ = $D_s
\delta\mbox{\boldmath $\beta$}$, Eq. (2) can be rewritten as
\begin{equation}
\delta(\Delta\tau_{BA}) = \frac{D}{cD_s} (1+z_d) [(\mbox{\boldmath $\theta$}_A 
- \mbox{\boldmath $\theta$}_B).\delta{\bf r}].
\end{equation}
Eq. (3) has two remarkable aspects. Firstly, {\it the time delay difference
does not depend on the lens model} (the distribution of mass in the lens).
This property is very important to do robust estimates with complex lenses
as the galaxy + cluster system in QSO 0957+561. Secondly, the time delay 
differences are only useful to map the projections of the relative positions 
of flares on the $\mbox{\boldmath $\theta$}_A - \mbox{\boldmath $\theta$}_B$ 
direction (see also Yonehara 1999).

In a cosmology with $\Omega_{\Lambda}$ = 0 and $\Omega_M$ = 1, the
observational parameters $z_d$ = 0.36, $z_s$ = 1.41 (the redshift of the
source) and $\mid\mbox{\boldmath $\theta$}_A - 
\mbox{\boldmath $\theta$}_B\mid$ = 6".1 lead to
\begin{equation}
[\delta(\Delta\tau_{BA})]({\rm days}) = 0.057 \delta y({\rm pc})  ,
\end{equation}
being $\delta y$ = $[(\mbox{\boldmath $\theta$}_A 
- \mbox{\boldmath $\theta$}_B).\delta{\bf r}]$/$\mid\mbox{\boldmath 
$\theta$}_A - \mbox{\boldmath $\theta$}_B\mid$. Here, the $y$-axis is 
parallel to $\mbox{\boldmath $\theta$}_A - \mbox{\boldmath $\theta$}_B$,
while the $x$-axis is perpendicular to this privileged direction of the
gravitational mirage. A more realistic cosmology (e.g., Wang et al. 2000) could
be suitable to obtain very refined estimates, but in this first approach to
the problem, our aim is not so ambitious. We wish to obtain some robust bound
on the size of the region associated with the flares inducing the prominent 
APO/TO events. The three  time delays claimed in Section 2 suggest a time delay
of 425 days for hypothetical signals emitted from the source centre (the 
average of all delays), as well as time delay differences of 
$\delta(\Delta\tau_{BA})$ = $-$ 8.0 $\pm$ 0.6 days (APO main events), 
$\delta(\Delta\tau_{BA})$ = $+$ 7.0 $\pm$ 1.9 days (APO secondary events) and 
$\delta(\Delta\tau_{BA})$ = 0 $\pm$ 4 days (TO events). From Eq. (4), it is 
easy to derive the projected relative positions ($\delta y$): $-$ 140 $\pm$ 10,
$+$ 123 $\pm$ 33 and 0 $\pm$ 70 pc. We ignore the components 
$\delta x$ corresponding to the three flares, however, the information on the 
components $\delta y$ is sufficient to reach important conclusions. Fig. 7 
presents the possible relative positions of the flares (shading rectangles). 
The rectangles corresponding to the two $g$-band flares are depicted in the
left panel, while the rectangle drawn in the right panel is associated with the
$R$-band flare. A circle with a radius of 90 pc is also shown in the left panel
of Fig. 7. {\it A region of flares with outer radius less than 90 pc is in 
disagreement with all possible relative positions of the g-band flares, and the
two far flares were probably generated at distances of 100-200 pc from the 
central black hole. This result indicates that at least two flares are related 
to a stellar environment}. On the other hand, the $R$-band flare might be
produced far from the central engine, and thus it could be also related to the
stellar ring. However, we cannot decide on this issue. 

\begin{figure}
\psfig{figure=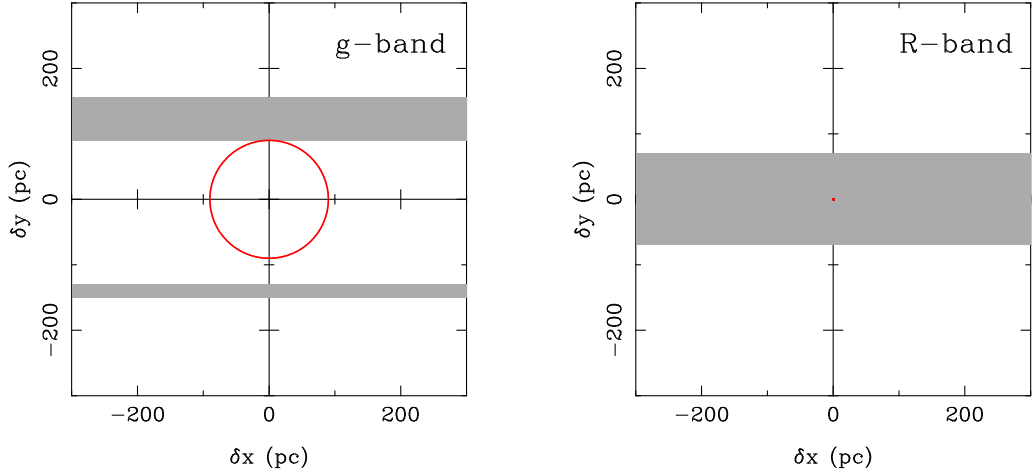,width=15cm,angle=-90}  
\caption{Flares around the centre of a standard source. Left panel: we can see 
the possible relative positions of the $g$-band flares associated with the 
prominent APO events (shading rectangles) together with a circle at 90 pc from 
the centre lodging a supermassive black hole. Right panel: the shading
rectangle represents the possible relative positions of the $R$-band flare
which is responsible for the TO events.}
\label{Fig. 7}
\end{figure}

\section{Conclusions and discussion}

Using the best available optical light curves of Q0957+561A,B, which contain
three pairs of twin events with significant signal-to-noise ratio ($S/N \geq$ 
2.5), we found a clear evidence for three different time delays. As far
as we know, this is the first detection of multiple delays in a double-imaged
QSO. Our analysis based on data of a lens system sampled during many years
could be extended to other lens systems in a near future.

We concentrated on relatively violent events with an amplitude of $\sim$ 100
mmag (a fluctuation of $\Delta m \sim$ 0.1 mag represents a relative fluctuation
in flux of $\Delta F/F_{back} \sim$ 10 per cent) and a duration of $\sim$ 
100-300 days. If these events are originated inside a standard hybrid source
(accretion disk and circumnuclear stellar region), the observed time delay 
multiplicity suggests that they do not come from a common zone in the source 
(Yonehara 1999). So, the studied events may be related to violent and local 
physical phenomena (flares). Accepting the hypothesis of events caused by 
flares in a standard hybrid source, in Sect. 3 we discussed on the size and 
nature of the region of flares. In particular, it is showed that two of the 
three flares must be generated at distances (from the central black hole) 
larger than 90 pc. Therefore, taking into account that the radius of a typical 
accretion disk is of 10$^{-2}$ pc, accretion disk instabilities or collisions 
of stars with the black hole + hot disk complex (e.g., Haardt, Maraschi \& 
Ghiselini 1994; Kawaguchi et al. 1998; Ayal, Livio \& Piran 2000) can be 
discarded as candidates to these two flares far away from the black hole. 
However, in principle, the observed flux variability may be due to either 
stellar collisions in dense clusters separated from the centre (e.g., 
Courvoisier, Paltani \& Walter 1996) or the violent evolution of stars in the 
circumnuclear stellar region (e.g., Aretxaga, Cid Fernandes \& Terlevich 1997).
A more complex scenario, involving star-gas encounters in a extended gaseous 
disk or bar, is also possible. 

We can roughly estimate the rest-frame $B$-band energy released in the flare 
related to the main events in the $g$-band APO data, and compare it with the
rest-frame $B$-band energy released in a typical event occuring in a given 
stellar scenario. To do the estimation/comparison we followed several steps. 
Firstly, from the light curve of the image A in the first season ($A95$), it is
deduced a $g$-band background of $m_{back}(g)$ = 17.08 mag. We transformed this
$g$-band background component to a $B$-band background contribution using the 
$g-r$ color and the equations of Kent (1985). The $B$ magnitude can be then 
converted to a monochromatic flux (4400 \AA) using standard laws (e.g., Allen 
1973; Henden and Kaitchuck 1990; L\'ena, Lebrun \& Mignard 1998). The relevant
flux, $F[\lambda 4400(1+z_s)]$, may be found by multiplying the monochromatic 
flux $F(\lambda 4400)$ by the $k$-correction $(1 + z_s)^{0.5}$ (e.g., Schmidt
\& Green 1983). Thus, $F_{back}[\lambda 4400(1+z_s)] \approx$ 10$^{-15}$ erg 
cm$^{-2}$ s$^{-1}$ \AA$^{-1}$. Secondly, using the luminosity distance $D_L$, 
one obtains: $\tau_{back} L_{back}(\lambda 4400) = 4\pi D_L^2 F_{back}[\lambda 
4400(1+z_s)]$, where $L_{back}(\lambda 4400)$ is the background intrinsic 
luminosity (at 4400 \AA) and $\tau_{back}$ is the extinction-magnification 
factor. This last factor is due to both the magnification of the light from 
image A by the lens, and the extinction by dust in the lens galaxy, the Milky 
Way and so on. The magnification can be easily obtained, whereas it is 
difficult to estimate the total extinction. In a cosmology with 
$\Omega_{\Lambda}$ = 0, $\Omega_M$ = 1 and $H$ = 100 $h$ km s$^{-1}$ 
Mpc$^{-1}$, we inferred $h^2 \tau_{back} L_{back}(\lambda 4400) \approx$ 3 
10$^{42}$ erg s$^{-1}$ \AA$^{-1}$. Thirdly, the observed event has a duration 
of $\Delta t$ = 100 days and an effective (top hat) amplitude of $\Delta m(g)$ 
= $-$ 0.07 mag. We assumed that $\Delta m(B) \approx$ $-$ 0.07 mag and found an 
intrinsic luminosity of the associated flare given by $h^2 \tau_{flare} 
L_{flare}(\lambda 4400) \approx$ 2 10$^{41}$ erg s$^{-1}$ \AA$^{-1}$. Taking 
$h^2 \tau_{flare} \sim$ 1, a time scale of $\Delta t_{flare} = \Delta t/(1 + 
z_s) \approx$ 40 days, and a width of the filter's bandpass of 1000 \AA, the 
rest-frame $B$-band energy released in the flare will be $E_{flare}(B) \sim 
10^{51}$ erg. Finally, we compared the result on $E_{flare}(B)$ with the
energy released in a supernova explosion and in a head-on stellar collision. 
The mean rest-frame $B$-band energy released in a SN explosion is of $E_{SN}(B)
\approx$ 0.5 10$^{51}$ erg (Aretxaga, Cid Fernandes \& Terlevich 1997). On the 
other hand, the energy released in a collision of two solar-mass stars with a 
velocity parameter $\beta = v/c$ will be of $E_{COL}(B) < E_{COL} \approx 
M_{\odot} c^2 \beta^2 \approx 10^{54}\beta^2$ erg (Courvoisier, Paltani \& 
Walter 1996). Therefore, the value of $E_{flare}(B)$ agrees with the energy 
released in a supernova explosion and in a relativistic stellar collision 
($\beta \approx$ 0.1), and a stellar collision at moderate velocities ($\beta 
\approx 10^{-3}-10^{-2}$) cannot produce so high energy in the rest-frame $B$ 
band.   

For Seyfert nuclei, the isolated SN events have a characteristic time-scale of
$\approx$ 300 days. However, the evolution of SNs in QSOs can be very much
faster, with time-scales of $\approx$ 10 days (see Aretxaga, Cid Fernandes \& 
Terlevich 1997). So, the time-scale and the energy corresponding to the 
$g$-band flares agree with the expected ones in a starburst scenario. On the 
other hand, for QSO 0957+561, a picture including only a starburst nucleus
or ring has one difficulty. Kawaguchi et al. (1998) claimed that the measured
slope of the first-order structure function is in clear disagreement with the
pure starburst model. However, as it was remarked by Kawaguchi et al. (1998),
the assumption of either a realistic shape of each SN event or a hybrid 
scenario could conciliate the existence of supernova explosions and the 
observed structure function. Moreover, the observational sampling was not taken
into account in the simulations, and the observed structure function included
observational noise, which was absent in model calculations. 
  
Very recently, Collier (2001) found evidence for accretion disk reprocessing in
QSO 0957+561. He compared the two events depicted in the top panel of Fig. 1 
(the $g$-band APO main event and the $r$-band APO main event in $B96$) and 
obtained that the event in the $r$ band lags the event in the $g$ band by 
$\sim$ 1 day. This non-zero chromatic lag seems to support a reverberation 
within an accretion disk, and therefore, flares originated at points very close
to the black hole. The result disagrees with our suggestion about the origin of
the flare related to the APO main events in the $g$ band, but with regard to 
this discrepancy, we must do a remark. The detection of a non-zero lag is only 
based in an interpolated cross-correlation method. Another discrete technique 
($ZDCF$) led to an 1 $\sigma$ interval including a lag equal to zero. In any 
case, we can conciliate the claim by Collier (2001) and the measured delays in 
a simple way. The $g$-band APO main events could come from a flare in the hot 
disk, with the other two flares (which generated the $g$-band APO secondary 
events and the $R$-band TO events) having a stellar origin and typical time 
delay differences of $+$ 8 and $+$ 15 days. In this new scheme, the projected 
(on a direction joining both images of the QSO) relative positions of the two 
stellar flares are $\delta y$ = $+$ 140 pc and $\delta y$ = $+$ 263 pc. 
However, the solution to the possible conflict is not very comfortable. As the 
direction defined by the vector $\mbox{\boldmath $\theta$}_A - \mbox{\boldmath 
$\theta$}_B$ is not a privileged one in relation to the physics of the source 
(obviously, it is a privileged direction of the gravitational mirage), there is
an apparent bias in the 2D distribution of stellar flares: the two flares are 
produced in the semicircle defined by $\mbox{\boldmath $\theta$}_A - 
\mbox{\boldmath $\theta$}_B$. Of course the statistic is very poor with only a
few flares, and we must analyze new monitoring campaigns to find more flares
and a reliable estimate of the time delay associated with hypothetical signals
from the centre of the source. 

\section*{Acknowledgments}

I am grateful to Atsunori Yonehara and Joachim Wambsganss for interesting 
discussions during the GLITP Workshop held at the Instituto de Astrofisica de 
Canarias (IAC). I also acknowledge the anonymous referee for criticisms that 
helped improve the paper. This work was supported by Universidad de Cantabria 
funds, the DGESIC (Spain) grant PB97-0220-C02 and the Spanish Department of
Science and Technology grant AYA2001-1647-C02.

\bsp

\end{document}